%% file: paper.tex
\documentclass[submission,copyright,creativecommons]{eptcs}
\providecommand{\event}{GANDALF 2020} 
\usepackage{breakurl}             
\usepackage{underscore}           
\usepackage{graphicx}
\usepackage[ruled]{algorithm}
\usepackage[noend]{algpseudocode}
\usepackage{xcolor}
\usepackage{framed}
\usepackage{amsfonts}
\usepackage[english]{babel}
\usepackage{amsmath}
\usepackage{float}
\usepackage{wrapfig}

\title{Symbolic Execution + Model Counting\\+ Entropy Maximization = Automatic Search Synthesis}
\author{Mara Downing
\institute{Harvey Mudd College\\ Claremont, CA, USA}
\email{mdowning@g.hmc.edu}
\and
Abtin Molavi
\institute{Harvey Mudd College\\ Claremont, CA, USA}
\email{amolavi@g.hmc.edu}
\and
Lucas Bang
\institute{Harvey Mudd College\\ Claremont, CA, USA}
\email{bang@cs.hmc.edu}
}
\def\titlerunning{Automatic Search Synthesis}
\def\authorrunning{Downing, Molavi, Bang}
\begin{document}
\maketitle

\begin{abstract}
We present a method of automatically synthesizing steps to solve search problems. Given a specification of a search problem, our approach uses \emph{symbolic execution} to analyze the specification in order to extract a set of constraints which model the problem. These constraints are used in a process called \emph{model counting}, which is leveraged to compute probability distributions relating search steps to predicates about an unknown target. The probability distribution functions determine an \emph{information gain objective function} based on Shannon entropy, which, when maximized, yields the next optimal step of the search. We prove that our algorithm converges to a correct solution, and discuss computational complexity issues. We implemented a domain specific language in which to write search problem specifications, enabling our static analysis phase. Our experiments demonstrate the effectiveness of our approach on a set of search problem case studies inspired by the domains of software security, computational geometry, AI for games, and user preference ranking.
\end{abstract}

\section{Introduction}
\label{sec:intro}

Searching is a fundamental problem in computer science. For example, classic search algorithms taught in introductory programming classes include linear search through unordered linked lists and binary search in sorted arrays. More complex search problems include object localization via bounding-box methods within an image (say, in computer vision)\cite{ORourke2004FindingME}, finding the optimal plays in an interactive puzzle game\cite{KlimosK15}, determining customer preferences via interactive iterative ranking\cite{PuKS17}, or discovering a software security vulnerability\cite{DBLP:conf/ccs/KopfB07}, to name a few. 

We observe that \textit{specifying} a search problem is almost always easier than \textit{solving} it. For instance, it is simple to write a function that checks if two rectangles overlap. However, it is more difficult to write a program that adaptively adjusts the size and location of a rectangular window until it exactly matches an unknown target rectangle. For this example, our system allows a user to write code implementing the bounding-box containment \textit{check}, and we then synthesize the online-optimal rectangular window \textit{search} to find a bounding box. At a higher level, one may view the synthesis of a search solution as a game between a searching algorithm and an oracle that reveals partial information about a search query at each step. Our approach provides a general technique, or \textit{meta-search} algorithm that solves this form of game when provided with a programmatic description of the search problem and access to the oracle.

In this paper we describe a general framework that, given a specification of a search problem, synthesizes optimal adaptive online search steps, thereby solving the specified search problem. Our framework is adaptive, in the sense that it uses information learned from previous search queries to inform future search steps. It is online in the sense that it produces search steps one at a time in response to a search step outcome. Our approach allows a user to specify a search problem as an imperative program. We perform symbolic execution on that search problem specification and use the resulting constraints in a model counting procedure to automatically generate an information-gain objective function which is maximized to synthesize search steps. This paper makes the following contributions:

\begin{itemize}

\item We define a \textit{meta-search} algorithm (Section~\ref{sec:algorithm}) that takes in the specification of a search problem and synthesizes adaptive online-optimal search steps to solve the problem.

\item We prove convergence and correctness of our algorithm (Section~\ref{sec:conv-correct}). 

\item We empirically validate our approach on a set of search problems (Section~\ref{sec:experiments}), showing applicability to domains like security exploit discovery, AI for games, and geometric object localization.

\end{itemize}

\section{Background and Overview}
\label{sec:overview}

We give an overview of our approach, including the definitions for our model of search problems, the steps of automatically solving a search problem, and follow with examples.

\subsection{Components of a Search Problem}

A search problem $P$ comprises target, query, and outcome sets and  query evaluation function: $\langle T, Q, O, E \rangle$. 

\vskip 0.5mm

\noindent \textbf{Target.} The solution to a search problem is to discover a \emph{target} value $t^*$ among a finite set of all potential target values $T$. For instance, the target might be an unknown integer in a range.

\vskip 0.5mm

\noindent \textbf{Queries.} A search makes queries $q^*$ from among a set of queries $Q$. E.g. $Q$ might be a range of integers. 

\vskip 0.5mm

\noindent \textbf{Search Problem Specification.} The specification expresses the relationship that holds between targets and queries. In our setting, it is encoded as an imperative function. E.g. \texttt{S(q,t) = return t < q}. 

\vskip 0.5mm

\noindent \textbf{Search Problem Instance.} Instantiated by plugging in a specific $t^*$, not revealed to the search procedure.

\vskip 0.5mm

\noindent \textbf{Evaluation.} A query is evaluated according to the spec, where the spec is instantiated with $q^*$ and a value of $t^*$ that is unknown to the search algorithm. We write $E(q^*, t^*)$ for the evaluation.

\vskip 0.5mm

\noindent \textbf{Outcomes.} After each query, the outcome, $o$, is revealed, from among a set of possible outcomes $O$.

\vskip 0.5mm

\noindent \textbf{Adaptive.} An adaptive search maintains knowledge $\kappa$ about the target $t^*$, learned via earlier queries, using $\kappa(t^*)$ to make later queries.

\vskip 0.5mm

\noindent \textbf{Online.} An online search decides on subsequent queries one at a time upon receiving each outcome.

\begin{figure}
  \centering
  \includegraphics[scale=0.7]{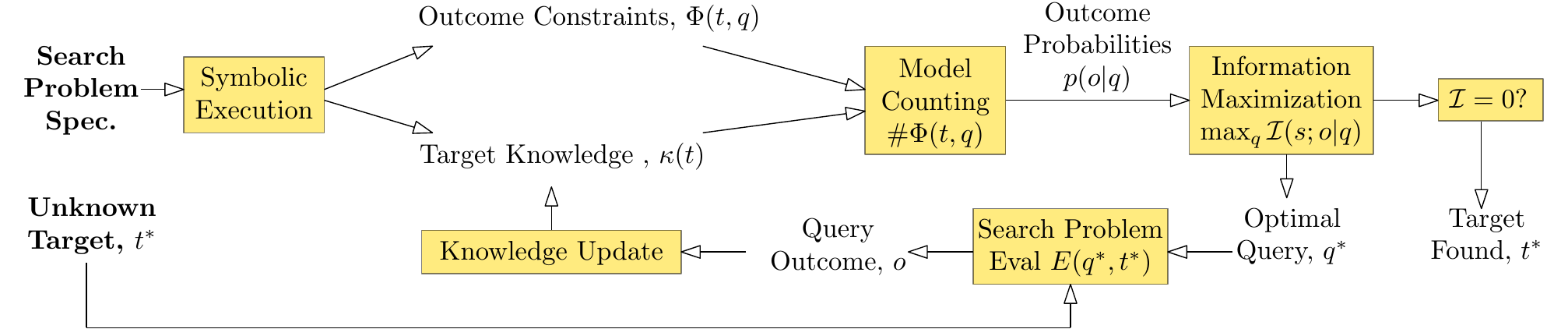}
  \caption{Overall strategy synthesis approach..}
  \label{fig:system-diagram}
\end{figure}

\subsection{Solution Synthesis Steps and Components}

We now describe our solution technique. The reader may find it helpful to refer to Figure~\ref{fig:system-diagram}.


\vskip 0.5mm

\noindent \textbf{1. System input.} Source code for a search problem specification is provided which defines (a) what are the possible target values and (b) the relationship between search queries, the target value, and a query outcome. We assume this relationship is specified by a deterministic program. (Our domain specific language for describing search problems is given in Section~\ref{sec:dsl}.) 

\vskip 0.5mm

\noindent \textbf{2. Static Specification Analysis}. 
We extract a symbolic model of the search problem from a specification written in  our search problem specification language. The symbolic model comes in the form of a set of outcome constraints using symbolic execution. Symbolic execution~\cite{King:1976:SEP:360248.360252} 
is a popular static code analysis technique by which a program is executed on \emph{symbolic} (as opposed to concrete) input values that represent all concrete values. In the limit, symbolic execution explores all feasible paths of execution. 

Symbolically executing a program yields a set of \emph{path
constraints} $\Psi = \{\psi_1, \psi_2, \ldots, \psi_m\}$. Each~$\psi_i$ is
a conjunction of constraints on the symbolic inputs that characterize all
concrete inputs that would cause a path to be followed. All the $\psi_i$'s
are disjoint. Whenever symbolic execution hits a branch condition $c$, both branches are explored and the constraint
is updated: $\psi \leftarrow \psi \wedge c$ in the \emph{true} branch
and $\psi \leftarrow \psi \wedge \neg c$ in the \emph{false} branch.  
Path constraint
satisfiability is checked using constraint solvers such
as Z3~\cite{DeMoura:2008:ZES:1792734.1792766}.
If a path constraint is unsatisfiable, that path pruned from the symbolic exploration. 

We treat the target value $t$ and the input query $q$ as symbolic and we associate each path constraint with the corresponding search query evaluation outcome. Thus, during symbolic execution, we track the return values of functions that implement the search problem specification and so each $\psi_i$ is associated with a concrete outcome $o_i$. Path constraints which result in the same outcome are combined using disjunction to produce the \emph{outcome constraints}. Symbolic execution of a search problem specification returns a set of constraints $\Phi = \{\phi_1, \phi_2, \ldots, \phi_n \}$, one per outcome, which are logical formulas encoding the relationship between input queries and target values corresponding to that outcome.

\vskip 0.5mm

\noindent \textbf{3. Probabilities via Model Counting.} We seek to compute probability distributions, $p(o|Q=q^*)$, the probability that a query $q$ will result in outcome $o$. 
It is sufficient to compute the number of targets $t$ that satisfy both the search algorithm's current knowledge about the target, $\kappa(t)$, and each outcome constraint $\phi_o$, all as a function of $q$, denoted $\#[\phi_{o} \land \kappa(t)](q)$.  Counting solutions to constraints is a well-studied problem, known as \emph{model counting}, and various tools exist for accomplishing this task for constraints over various types (e.g. integers, strings, booleans)~\cite{ABB15,DeLoera20041273,LSS14,Barvinok:1994:PTA:187096.187093,website:barvinok}. The probability of an outcome given a query is easily computed using these model counts: $p(o|Q=q^*) = \#[\phi_o \land \kappa(t)](q) / \#\kappa(t)$, where $\# f $ is the number of satisfying solutions to a constraint formula $f$ and $\# f(v) $ is the number of satisfying solutions to a constraint formula $f$ as a function of some variable $v$.

\vskip 0.5mm

\noindent \textbf{4. Information Gain Maximization.} Using $p(o|Q=q^*)$, we compute the conditional mutual Shannon information~\cite{shannon48,Cover2006} between an experiment outcome $O$ and target $T$, given search query $Q$, denoted $\mathcal{I}(T;O | Q = q^*)$. $\mathcal{I}$ is the amount of information that the search algorithm expects to gain about $t^*$ by receiving outcome $o$ after making query $q^*$. We compute
$\mathcal{I}(T;O | Q = q^*) = \mathcal{H}(q^*)$
where $\mathcal{H}$ is the Shannon entropy. A full explanation is given in Section~\ref{sec:objective_function}. We maximize $\mathcal{I}$ to find query $q^*$ with the highest expected information gain about target $t^*$: $q^* = \arg \max_{q^* \in Q} \mathcal{I}(T;O|Q = q^*)$.

\vskip 0.5mm


\noindent \textbf{5. Evaluate the Query and Update Knowledge.} Query $q^*$ is evaluated, producing outcome $o$ according to the search problem specification. Since each outcome $o$ is associated with a constraint on $q$ and $t$, the search algorithm can update the knowledge about the target as the conjunction of the current knowledge $\kappa(t)$ with the corresponding observation constraint, replacing the query variable $q$ with the query that was evaluated, $q^*$, denoted $\phi_o[q \mapsto q^*]$:
$\kappa(t) \leftarrow \kappa(t) \land \phi_o[q \mapsto q^*]$.

\vskip 0.5mm

\noindent \textbf{6. Repeat Until No Information Gain is Possible.} This process repeats until a fixed-point is achieved in which there are no more queries that can result in information gain. In Section~\ref{sec:filter} we describe the stopping criteria, and use it to guarantee convergence to $t^*$ in finitely many steps.

\vskip 1mm

\label{sec:intervalsearch}

\noindent \textbf{EXAMPLE: Interval Searching.} We now walk through the steps just outlined for a simple search problem which touches on the main points of our approach. 

\begin{figure}
  \centering
  \begin{minipage}{65mm}
  \begin{verbatim}
  function evaluate(q, t)
    if t < q[0]
      return "Low"
    else if q[0] <= t <= q[1]
      return "Middle"
    else
      return "High"
  \end{verbatim}
  \end{minipage}
  \begin{minipage}{65mm}
 \includegraphics[scale=0.30]{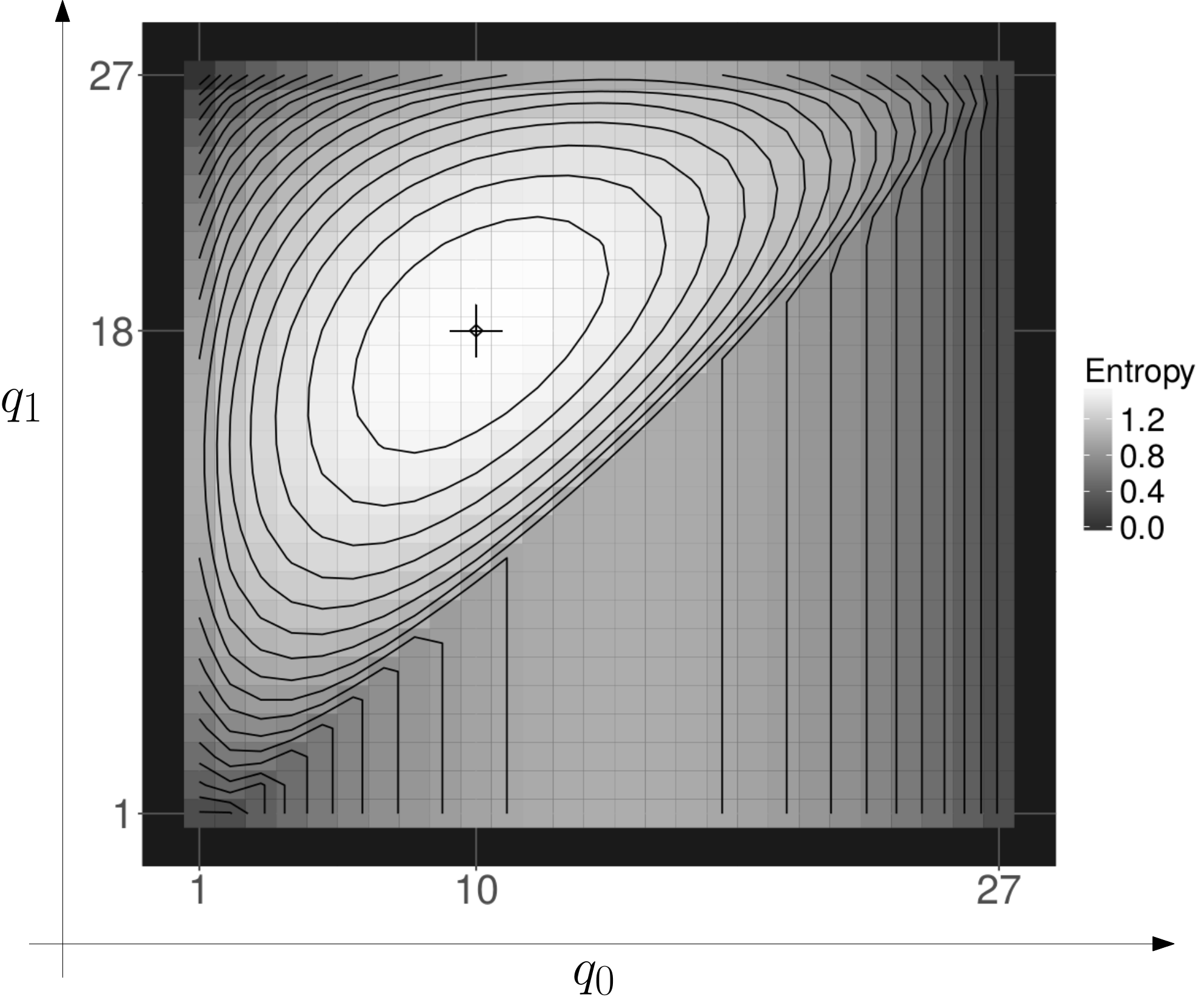}
  \end{minipage}

%

  \caption{Left: Code for \textsc{Low-Middle-High} search. Right: Contour plot of information gain $\mathcal{I}$, for \textsc{Low-Middle-High} search where $1 \leq t \leq 27$. Maximum occurs at $(10,18)$, indicated by the cross-hair.}
  \label{fig:lmh-code}
\end{figure}

\vskip 0.5mm
\noindent \textbf{Search Problem Specification.} 
Consider a search problem that we will call \textsc{Low-Middle-High} in which the target is an unknown integer within a known range: $1 \leq t \leq 27$. A query is a pair of integers $q = (q_0, q_1)$, which we interpret as a lower and upper bound of an integer interval. The search algorithm is informed of the value of the target $t$ \textit{relative to the query}. That is, the outcome of a query is determined according to the \texttt{experiment} function in the pseudocode of 
Figure~\ref{fig:lmh-code}: 
``Low'' if $t < q_0$, 
``Middle'' if $q_0 \leq t \leq q_1$,
and ``High'' if neither of those cases apply.

\vskip 0.5mm
\noindent \textbf{Intuitive search strategy.} 
One might reason that the best strategy solving this search problem is to choose $(q_0,q_1)$ at each round such that the search space is cut into equal thirds every time. Indeed, this is the optimal strategy. That is, this \textit{ternary searching} strategy minimizes the expected number of queries needed to find a given target. In fact, this strategy also guarantees that the target will be found within $\log_3 n$ steps when there are $n$ targets. We now walk through how our approach synthesizes this solution. 

\vskip 0.5mm

\noindent \textbf{Search parameters.} For this problem, target set is $T = \{s : 1 \leq t \leq 27\}$, query set is $Q = \mathbb{Z} \times \mathbb{Z}$, the set of outcomes is $O = \{$ `Low', `Middle', `High' $\}$ which we abbreviate to $O = \{L, M, H\}$.

\vskip 0.5mm
\noindent \textbf{Static analysis.} 
Performing symbolic execution of the specification results in the following \emph{outcome constraints}. For this simple example, one can easily see how the constraints correspond to the function which implements the specification: \hspace{7mm}$\phi_L  =  t < q_0, \ \ \ \ \ \ \phi_M  =  q_0 \leq t \leq q_1, \ \ \ \ \ \ \phi_H  =  t > q_1$



\vskip 0.5mm

\noindent \textbf{Conditional outcome probabilities via model counting.} 
We now wish to know the probabilities of each outcome as a function of the query. Consider constraint $\phi_L$. There are two non-trivial cases. If $q_0 \geq 27$ there are $27$ secret values consistent with the knowledge $\kappa(t) = 1 \leq t \leq 27$. On the other hand, if $1 \leq q_0 < 27$, then there are $q_0 - 1$ solutions for $\kappa(t)$, namely, $s \in \{1, \ldots, q_0 - 1\}$. For any other values of $q_0$, $\kappa(t)$ is unsatisfiable and so there 0 solutions. Reasoning about $\#\phi_{H}(q_0,q_1)$ is symmetrically similar, and $\#\phi_{M}(q_0,q_1)$ is slightly more complicated, giving us the three piecewise counting functions shown below. These counting functions are produced automatically using model counting tools. In this case, we used the Barvinok model counting library~\cite{website:barvinok}. With these counting functions, we can easily compute the probability of each outcome, $o$, conditioned on the query: $p(o|q) = \#[\phi_{o} \land \kappa(t)](q) / \#\kappa(t)$. 


\begin{minipage}[t]{0.4\textwidth}
\vspace{0mm}
{\footnotesize
$
\#\phi_{L}(q_0,q_1) = 
\begin{cases} 
27      & q_0 \geq 27 \\
q_0 - 1 & 1 \leq q_0 < 27 \\
0       & \text{otherwise}  
\end{cases}
$\\ \\

$
\#\phi_{H}(q_0,q_1) = 
\begin{cases} 
27       & q_1 < 0 \\
27 - q_1 & 0 \leq q_1 < 27 \\
0        & \text{otherwise}  
\end{cases}
$\\ \\
}
\end{minipage}
\begin{minipage}[t]{0.5\textwidth}
\vspace{0mm}
{\footnotesize
$
\#\phi_{M}(q_0,q_1) = 
\begin{cases} 
27              & q_0 \leq 1 \land q_1 > 27 \\
28 - q_0        & 1 < q_0 \leq 27 \land q_1 > 27 \\ 
q_1 - q_0 + 1   & q_0 < 1 \land q_0 \leq q_1 \leq 27 \\ 
q_1             & q_0 \leq 1 \land 0 < q_1 \leq 27 \\ 
0               & \text{otherwise}  
\end{cases}
$ \\
}
\end{minipage}

\vskip 0.5mm

\noindent \textbf{Optimal query via information maximization.} 
From the outcome probabilities we compute the information gain using $p(L|q_0,q_1)$, $p(M|q_0,q_1)$, and $p(H|q_0,q_1)$ by plugging the appropriate probability expressions. See Section~\ref{sec:objective_function} for details of this computation. A contour plot of the \textsc{Low-Middle-High} 
information gain objective function 
$\mathcal{I}$ as a function of 
$(q_0,q_1)$ is given in 
Figure~\ref{fig:lmh-code}. The point in this space that maximizes $\mathcal{H}$ is the query 
$q^* = \arg \max \mathcal{H}(o|(q_0,q_1)) = (10,18).$
\noindent This aligns with our earlier intuition that the best strategy is to split the search space in thirds during each round. 

\vskip 0.5mm

\noindent\textbf{Knowledge update based on query outcome.} The search algorithm then learns the outcome $o$ for query $q^*$ according to the spec: $o = R(q^*, s)$. In our example, one can learn that $1 \leq t \leq 9$ if the outcome is ``Low'', $10 \leq t \leq 18$ if the outcome is ``Middle'', or $19 \leq t \leq 27$ if the outcome is ``High''. In our running example, if the outcome $o = $ ``Low'' after playing query $q^* = (10,18)$, the update is: 

\vskip 0.5mm

\ \ \ \ \ \ \ \ \ $\kappa(t) \leftarrow  1 \leq t \leq 27 \land \phi_L[(q_0,q_1) \mapsto (10,18)] \equiv  1 \leq t \leq 27 \land t < 10 \equiv  1 \leq t \leq 9$


\vskip 0.5mm

We have demonstrated that entropy maximization based on the constraints generated by static analysis of the spec produces the first step of the optimal \emph{ternary search}. This is a more general principle which applies to more complex search specifications, as seen in our experimental results (Section~\ref{sec:experiments}). 

\vskip 0.5mm

\noindent \textbf{Query until information is exhausted.} Given updated $\kappa(t)$, the process repeats starting with model counting. Note that the static code analysis phase is not repeated, as the symbolic execution constraints in conjunction with the new knowledge sufficiently capture the behavior of the search problem. Supposing that the knowledge is updated as described in the previous step, the next round of query synthesis using model counting and entropy maximization results in $q^* = (4,6)$ (another step of ternary search). This continues until there are no queries with positive information gain, i.e. $\forall q [ \mathcal{I}(S; O| Q = q) = 0]$.

\section{Search Problem Online Adaptive Solution Synthesis}
\label{sec:synthesis}

In this section we give details behind our search synthesis procedure: the objective function based on Shannon entropy, reducing the query set search space at every iteration, the algorithm, and a proof sketch of termination and correctness.

\subsection{Objective Function for Information Gain}
\vskip 0.5mm
\label{sec:objective_function}
Here we derive an objective function to measure the amount of information any search algorithm expects to gain for a query $q^*$. By a slight abuse of notation, let $T$, $Q$, and $O$ be random variables representing the targets, queries, and outcomes. We use entropy-based metrics from the theory of quantitative information flow~\cite{Smi09}. For every query $q^*$ in Q, an entropy value can be calculated from the probability that that query produces each possible outcome. These probabilities are calculated using \textit{model counting} by computing the number of ways query $q^*$ can produce outcome $o$ (in $O$) and dividing by the total number of possible targets, $|T|$. Each query is associated with a list of probabilities equal to the total number of outcomes possible in the game, ${p(o_1 | q^*), p(o_2 | q^*), \cdots, p(0_n | q^*)}$. \textit{Shannon Entropy} for that query is then:

\begin{equation}
\mathcal{H}(q^*) = 
-\sum_{o_i \in \mathbb{O}} p(o_i | Q=q^*) \log_2 {p(o_i | Q=q^*)}
\label{eq:condentropy}
\end{equation}

\subsection{The Worthwhile Query Set}
\label{sec:filter}

\noindent \textbf{Motivation.} At each step, the search must find $q^*$ among a possible set of $Q$ values so that evaluating $E(q, t)$ will gain information. That is, the search algorithm must choose a $q^*$ that accomplishes two complementary objectives: (1) some outcome $o$ is consistent with $\kappa(t)$ and some $\varphi_o(t,q)$ is satisfiable (i.e. some outcome is possible), and (2) the algorithm does not learn a redundant constraint. We define a predicate $W : Q \rightarrow \{T,F\}$, where $W(q)$ is true if $\mathcal{I}(s;q|o) > 0$ and false otherwise. That is, $W(q)$ is true if the search algorithm expects to gain information by making query $q$. When this occurs, we say that $q$ is a \emph{worthwhile} query. Now we provide a definition of $W(q)$ and prove that it fully characterizes the set of informative queries. In addition, $W(q)$ provides a sound and complete stopping condition for our algorithm. If $W(q) \equiv \mathsf{false}$, the algorithm has gained as much information about $t^*$ as possible.

\vskip 2mm

\noindent \textbf{Short Example.} Recall the interval searching example (Section~\ref{sec:overview}). Suppose $\kappa(t) \equiv 10 \leq t \leq 18$. We have the outcome constraints 
$\phi_L \equiv  t \leq q_0 , 
 \phi_M \equiv   q_0 < t \leq q_1 , 
 \phi_H \equiv   t > q_0 \land s > q_1 $.
Now consider, should the search algorithm ever try a query like $(q_0, q_1) = (3,7)$? One can reason that if $\phi_L$ where to hold for that query, then $t < 3 \land 10 \leq t \leq 18$ but this is not possible. Likewise, for $\phi_M$, $3 < t \leq 7 \land 10 \leq t \leq 18$ is impossible. However, for $\phi_H$, $t > 3 \land t > 7 \land 10 \leq t \leq 18$ is possible, but $ 10 \leq t \leq 18 \Rightarrow t > 3 \land t > 7$, and so this would be redundant information; $(3,7)$ is a useless query. One might begin to imagine that the components of the set of informative queries is simply equal to the set corresponding to $\kappa(t)$. However, this is not the case. For instance, perhaps $Q$ is the set of possible \textit{indices} of a target located in an array \texttt{A}, and array values are positive values less than 100, but $0 \leq q <$ \texttt{length(A)}. We now give a generic way to compute a symbolic representation of worthwhile queries for a current search step.

\vskip 2mm

\noindent \textbf{Query Filtering Predicate.} The purpose of $W(q)$ is to filter out queries that are useless because they make all outcome constraints either impossible or redundant. We define 
\begin{equation}
\label{eq:filter}
W(\hat q) 
\equiv 
\bigvee_{o \in O} 
(\exists \ t [\kappa(t) \land \phi_o[q \mapsto \hat q]]) 
\land (\exists \ t [\kappa(t) \land \neg \phi_o[q \mapsto \hat q]])
\end{equation}
\noindent where, for a particular outcome $o$, the left conjunct enforces that $\phi_o$ is satisfiable along with the current knowledge $\kappa(t)$, and the right conjunct ensures that $\phi_o$ is not already implied by $\kappa(t)$.

\vskip 2mm 

\noindent \textsc{Theorem 1.}
$\forall \hat q$ [$W(\hat q) \Leftrightarrow \mathcal{I}(T;O | Q = \hat q) > 0$]. That is, $\hat q$ satisfies the query filter predicate if and only if the algorithm expects to gain information by using query $\hat q$.


\vskip 2mm 

\noindent \textsc{Proof.}
Suppose $W(\hat q)$ for some $\hat q$. Then for some $o \in O$, for some $t_1$, $t_2$, we have that $t_1$ satisfies $\kappa(t) \land \phi_o[q \mapsto \hat q]$ and $t_2$ satisfies $\kappa(t) \land \neg \phi_o[q \mapsto \hat q]$. It cannot be the case that $t_1 = t_2$, or else $\phi_o(t, \hat q)$ would be simultaneously $\mathsf{true}$ and $\mathsf{false}$. Since $t_2 \models \kappa(t) \land \neg \phi_o[q \mapsto \hat q]$, there must be another $o'$ such that $t_2 \models \kappa \land \phi_{o'}[q \mapsto \hat q]$. Since $p(o|q^*) = {\# (\kappa(t) \land \phi_o)[q \mapsto \hat q]} / \# \kappa(t)$, then $0 < p(o | \hat q) \leq 1$. Thus, the probability mass for $p(o | \hat q)$ is not concentrated on a single outcome so $\mathcal{H}(O|Q) > 0$. Without belaboring routine calculation, we appeal to well know information theoretic inequalities~\cite{Cover2006}, to conclude $\mathcal{I}(T;O|Q = \hat q) > 0$. It is straightforward to reverse this argument to show the biconditional.

\vskip 2mm

\noindent \textbf{Computing the Worthwhile Query Set.} We now define the worthwhile query set $Q^* = \{ q : W(q) \}$. We observe that $\kappa(t)$ and $\Phi$, and therefore $W$, have a symbolic representation. Thus, we are able to represent very large query spaces by not maintaining a concrete set. Consequently, we may use any symbolic reasoning tool (like \textsc{Z3}\cite{DeMoura:2008:ZES:1792734.1792766} or 
\textsc{Barvinok}\cite{website:barvinok})to compute a representation of $Q^*$ by performing existential quantifier elimination (projection) on Equation~\ref{eq:filter}.

\subsection{Complexity Issues}
\label{sec:complexity}
\newtheorem{defOS}{Definition}

Here we observe that determining the optimal sequence of queries for a search problem in our setting is intractable. To gain some intuition, suppose that an optimal search for a given problem instance requires $m$ steps, and the optimal query sequence is 
$\vec{q} =  q^*_1,q^*_2, \ldots, q^*_m$. 
If we let $Q^*_0$ be the initial worthwhile query set, $\vec{q}$ is one sequence from among,  
$(Q^*_0)^{m}$ 
possible query sequences. Hence, intuitively it appears that an optimal offline solution would need to optimize over an exponentially sized search space. In this section, we make this intuition more concrete and prove that the optimal offline search problem solution as defined in this paper is NP-Hard.

\vskip 2mm

\noindent \textbf{Preliminaries.} Let a \textit{search procedure} be an algorithm that choose queries according to some selection function until no further information gain is possible. The query selection function is a map $f$ from $\mathcal{P}(T)$ to $Q$. At each step, the search procedure provides the query $f(K)$ where $K$ represents the subset of $T$ consistent with $\kappa(t)$. We formulate the following decision problem to capture the notion of an optimal search procedure.

\vspace{1mm}

\noindent \textsc{Definition (Optimal Search).} \ Given a search problem $\langle T, Q, O, E \rangle$, and integer $w$, does there exist a search procedure such that the expected number of queries before termination is $w$ or fewer? We will refer to the optimal search problem as \textsf{OS}.

\vskip 0.5mm

\noindent \textsc{Definition (Optimal Decision Tree).} Given a finite set of items $X$ and a set of tests $\mathcal{T}$, the optimal decision tree is a tree where the leaves are the elements of $X$ and internal nodes are Boolean tests $\tau_i$. A path from the root to some $x \in X$ defines a sequence of binary tests that uniquely determine $x$. The Optimal Decision Tree problem asks if there is a decision tree where the total path length is not greater than a given weight $w$.

\noindent \textsc{Definition (Identification Procedure).} An identification procedure is a binary decision tree such that all non-terminal nodes are identified with a test and all terminal nodes are associated with an object in $X$.

\vspace{2mm}

\noindent \textsc{Theorem.} 
Optimal Search is NP-Hard

\vspace{2mm}

\noindent \textsc{Proof.} 
We will show that \textsf{DT} $\prec_p $ \textsf{OS} where $DT$ is the \textit{decision tree} problem as defined by Hayfil and Rivest~\cite{HYAFIL197615}, where $\prec_p $ is the polynomial-time reducibility relation. (Their proof that \text{DT} is NP-Hard employs a reduction from \textsf{Exact-Cover-3} to \textsf{DT}.)


As our reduction will make clear, \textsf{DT} can be thought of as a special case of \textsf{OS}. 
Let $(\mathcal{T}, X, w)$ be an arbitrary \textsf{DT} instance. We will construct a corresponding instance of \textsf{OS}, $\langle T, Q, O, R, w' \rangle$ as follows. Let $T=X$, $Q=\mathcal{T}$, and $O = \{0, 1\}$ where $R(T) = T(s)$. Finally set $w' = w / |X|$. This is clearly a polynomial time reduction; the only real computation done is the arithmetic to find the value of $w'$.

If there exists some identification procedure with external path length $w$, then it can be converted into a search procedure of the appropriate expected query number by performing the same series of tests. The first query made is is the test at the root node of the identification procedure, the next query is the test that the identification procedure would perform upon the received response, and so on. This is compatible with our definition of a query selection function because each internal node corresponds to a unique current knowledge $\kappa(t)$. There is no further information gain possible when exactly one element of $T$ is consistent with $\kappa(t)$. Since there is a unique path to each element of $X$ in the identification procedure, and the total length of all paths is $w$, the expected path length is simply the average path length $w /|X|$.

The same conversion holds in the converse direction. If there exists some search procedure with an expected number of queries of $\frac{w}{|X|}$, then we can construct an identification procedure by choosing $T_r = f(T)$ (the query given no information) as the root test, $f(\{x \in T : \ T_r(x) \})$ as the left branch and $f(\{x \in T : \neg T_r(x) \}$ as the right branch, continuining recursively in this manner only one element of $T$ is consistent with the current knowledge. Since the expected length of a path on this tree is $w/ |X|$, the total path length is $w$. Consequently, optimal search in our setting is NP-Hard.

\vskip 0.5mm 
\noindent \textbf{Takeaway.} Since our problem is intractable in general, this justifies that our approach that proceeds in a greedy fashion, one step at a time, while possibly not generating the optimal solution, is a worthwhile heuristic. This transforms our problem from a single optimization problem over exponentially many possible sequences of queries to many optimization problems over $|Q|$ queries.

\subsection{Meta-Search Algorithm}
\label{sec:algorithm}
\label{sec:algorithms}
The algorithm for search problem solution synthesis and information gain (objective function) computation is given in this section. Combining everything that we have discussed in this paper, we can compactly describe our meta-search algorithm, or search query synthesizer. One may find it useful to refer back to Figure~\ref{fig:system-diagram} and Sections~\ref{sec:overview} and~\ref{sec:synthesis} to parse this algorithm.

\begin{algorithm}
\caption{\textsc{SynthesizeQueries} Input: search problem specification, $P = \langle T, Q, O, E \rangle$, target $t^*$. Output: queries $q^*$ to solve search problem. }
\label{alg:synthesize}
\begin{algorithmic}[1]
\Procedure{SynthesizeQueries($P$)}{}

\State $\langle \Phi , \kappa(s) \rangle \leftarrow$ \textsc{SymbolicExecution}$(E)$


\State $Q^* \leftarrow \{q : W(q)\}$ via Eq.~\ref{eq:filter}

\While{$ Q^* \not = \emptyset$}

\State $\mathcal{I}(T;O | q) \leftarrow $ \textsc{MutualInformation}$(\Phi , \kappa(t))$ \ \ \ (via Model Counting and Eq.~\ref{eq:condentropy})

\State $q^* \leftarrow \arg \max_{q \in Q^*} \mathcal{I}(T;O | q) $

\State $o \leftarrow  E(q^*,t^*)$

\State $\kappa(t) \leftarrow \kappa(t) \land \phi_o[q \mapsto q^*]$

\State $Q^* \leftarrow \{q : W(q)\}$ via Eq.~\ref{eq:filter}


\EndWhile

\EndProcedure
\end{algorithmic}
\end{algorithm} 

  
  



  



  

  

  

  

\subsection{Algorithm Convergence and Correctness}
\label{sec:conv-correct}
Here we argue that the \textsc{SynthesizeQueries} algorithm terminates. Furthermore, when \textsc{SynthesizeQueries} does terminate, it is not possible to gain any more information about the target, and so has provided a sequence of queries that solve the specified search problem. The proofs behind these cast the algorithms as a fixed-point computation which reduces $Q^*$ at every step until it is empty.

\vskip 0.5mm
\noindent \textsc{Theorem 2.} \textsc{SynthesizeQueries} (Algorithm~\ref{alg:synthesize}) terminates with a correct solution.

\vskip 0.5mm
\noindent \textsc{Proof.}
We rely on the use of $W(q)$ (Equation~\ref{eq:filter}) as the stopping condition of Algorithm~\ref{alg:synthesize}. Because $T$ is a finite set of targets, $Q^*$ is also a finite set, which can be observed by noticing that in Equation~\ref{eq:filter}, we are taking a finite disjunction across formulas that are existentially quantified over a finite set. Next, observe that $W(q)$ prevents the search algorithm from trying the same value of $q^*$ twice in the same search. To see why this is so, suppose that at some step, query $q^*$ is evaluated, resulting in outcome $o$. Then the knowledge will be updated as $\kappa'(t) \leftarrow \kappa(t) \land \phi_o[q \mapsto q^*]$. Suppose for a contradiction that in the next step $q^*$ is used again, which must result in the same outcome. Then considering the right hand conjunct of Equation~\ref{eq:filter}, we would be asking that $\exists t [\kappa'(t) \land \phi_o[q \mapsto q^*] \land \neg \phi_o[q \mapsto q^*]]$ which is not possible. Hence, regardless of what queries are produced by Algorithm~\ref{alg:synthesize}, it will never try the same query twice back to back. It is straightforward to extend this reasoning to see that the search algorithm will also not ever try the same two queries in any sequence, so long as it computes $Q^*$ via Eq.~\ref{eq:filter}. Thus, we certainly eliminate at least one query in every step, meaning that $Q^*$ eventually becomes empty. Furthermore, when $Q^*$ is empty, by \textsc{Theorem 1}, $\mathcal{I} = 0$ and so no more information can be gained about $t$. Thus, $\kappa(t)$ has become as constrained as possible. As we have not made any assumptions of optimal querying in this proof, these results hold regardless of how $q^*$ is chosen at each step.




\section{Implementation and Experiments}

We implemented the approach of Section~\ref{sec:overview} by implementing Algorithm~\ref{alg:synthesize} in Python and interfacing with \textsc{Z3} for constraint satisfiability checking during symbolic execution and \textsc{Barvinok} for model counting. 


\subsection{Numeric Computing }

\vskip 0.5mm
\noindent \textbf{Model Counting.} In our implementation, we used the parametric model counting software \textsc{Barvinok}, an implementation of Barvinok's polynomial-time integer lattice point enumeration algorithm. \textsc{Barvinok} represents a constraint $C$ on variables $(s,q)$ as symbolic polytopes $\mathcal{P} \subseteq \mathbb{R}^m$. Barvinok's algorithm generates a multivariate piecewise polynomial function whose domain is a disjunction of polytope chambers $\mathcal{Q} \subseteq \mathbb{R}^m$ represented by linear half-spaces in $\mathbb{R}^m$. ~\cite{Barvinok:1994:PTA:187096.187093,website:barvinok}. (See for example the model counting functions of Section~\ref{sec:overview}, Example 1.) The resulting piecewise polynomial evaluates to the number of assignments of integer values to $\mathcal{P}$ that lie in the interior of $\mathcal{Q}$.

\vskip 0.5mm
\noindent \textbf{Information gain Maximization}. To maximize $\mathcal{I}(T;O|Q = q^*)$, we make use of the polytope chambers given by \textsc{Barvinok} in the model counting process. We perform accept-reject sampling~\cite{accept-reject-sampling} sampling from the $\mathcal{Q}$ chambers, evaluating $\mathcal{I}$ for each sample, and returning $q^*$ with the largest information gain.



\subsection{Search Problem Specification Language}
\label{sec:dsl}

\begin{figure}[!htbp]
\begin{minipage}[t]{0.5\textwidth}
{\scriptsize
  $$
\begin{array}{rcl}
\mathsf{Program}  & :=  & \sf{List(Stmt)} \\
\sf{Stmt}         & :=  & \sf{List(Stmt)} \\
                  & |   & \sf{If(BoolExp, Stmt)} \\
                  & |   & \sf{IfElse(BoolExp, Stmt, Stmt)} \\
                  & |   & \sf{While(BoolExp, Stmt)} \\
                  & |   & \sf{Assign(Id, {Exp \ | \ List(Exp)})} \\
                  & |   & \sf{ArrayStore(Id, IntExp, Exp)} \\ 
                  & |   & \sf{Return(Exp)} \\
                  & |   & \sf{FunctionDefine(Id, List(Exp), Stmt)} \\
\sf{Exp}          & :=  & \sf{BoolExp \ | \ IntExp} \\
                  & |   & \sf{ArrayDeclare(Id, IntExp)} \\
                  & |   & \sf{ArrayAccess(Id, IntExp)} \\
                  & |   & \sf{FunctionCall(Id, List(Id))} \\
                  & |   & \sf{Length(Id)} \\
\end{array}
$$}
\end{minipage}
\begin{minipage}[t]{0.5\textwidth}
{\scriptsize
  $$
\begin{array}{rcl}
\mathsf{BoolExp}  & :=  & \sf{true \ | \ false} \\
                  & |   & \sf{And(BoolExp, BoolExp)} \\
                  & |   & \sf{Or(BoolExp, BoolExp)} \\
                  & |   & \sf{Not(BoolExp)} \\
                  & |   & \sf{Less(IntExp, IntExp)} \\
                  & |   & \sf{Equal(IntExp, IntExp)} \\
\sf{IntExp}       & :=  & \sf{IntConst} \\
                  & |   & \sf{Plus(IntExp, IntExp)} \\
                  & |   & \sf{Times(IntExp, IntExp)} \\
\sf{IntConst}     & :=  & c \in \mathbb{Z}
\end{array} 
$$}
\end{minipage}

\caption{Domain specific language abstract grammar for specifying search problems, supporting basic imperative constructs, Boolean and integer operations, arrays, and functions.}
\label{fig:dsl}
\end{figure}

Our approach relies on extracting a logical representation of the search problem from the specification. To facilitate this, we designed a small language for encoding search problem specifications that strikes a balance between analyzability and expressiveness. The language has features that we found necessary to express search problems but is simple enough that writing static analysis routines and interfacing with the Z3 constraint solver under the hood is straightforward. The interpreter of our language is written in Python and supports basic imperative programming features including Boolean and integer operations, control and iteration structures, functions, and arrays. See Figure~\ref{fig:dsl} for the abstract grammar. While we could possibly have used existing symbolic execution tools, we found that implementing our own compact spec language provided dexterity and agility in developing the overall system.

\subsection{Experimental Evaluation}
\label{sec:experiments}

\input{tables/data-tables.tex}

We conducted several case studies across a variety of problem domains. Our solver was able to synthesize queries for problems arising from logical reasoning puzzles, hidden state board games, software security exploit detection, user preference ranking, numeric searching, and geometric searching. We wrote specifications for each search problem in the domain specific language of Section~\ref{sec:dsl} with results shown in tables within this section. To collect data, we solved each game 10 times with different randomly chosen target values. We report the number of path conditions $|\Psi|$, the number of outcome constraints $|\Phi|$ after disjunctive merging (Section~\ref{sec:overview}), size of the query space $|Q|$, size of the secret search space $|S|$,  and averages of the symbolic execution time, game solving time, and number of rounds required to solve the game. We now give details of problems from each search problem domain that we explored. 

\subsubsection{Logical Reasoning Puzzles}

\noindent \textbf{Counterfeit Coin.} 
This is the classic counterfeit coin problem. There are $n$ coins that look identical, one of which is slightly lighter or heavier than the others. The player can place any number of coins on either side of a scale which tilts left, right, or balances. The goal is to discover which coin is the counterfeit with as few weighings as possible~\cite{coinproblem}.

\vskip 0.5mm
\noindent \textbf{Horse Race.} 
A player is attempting to discover the order in which 5 horses will finish a race, but you can only race 3 horses at a time and discover the order in which they finish.


\subsubsection{Security Exploit Synthesis}

We may think of static security analysis as a search for an exploit. Finding an exploit demonstrates the existence of a vulnerability. On the other hand, failure to synthesize the exploit on a repaired version of the code, while not guaranteeing security, provides some confidence that the vulnerability has been fixed. 

\noindent \textbf{Password Checker.} 
This models a security bug known as a segment oracle prefix attack\cite{BAP16}. In this example, a loop compares linear data structures (e.g. arrays or contiguous memory blocks) until a mismatch is found, in which case it returns \texttt{false}, and otherwise returns \texttt{true} if no mismatch is found. The amount of time that the loop runs leaks information about the length of the match, allowing an attacker to iteratively probe the system and discover secrets. By implementing the above logic in our DSL and running our meta-search algorithm, we are able to synthesize and therefore demonstrate the vulnerability. 

\noindent \textbf{Secured Password Checker.} 
We repaired the vulnerability by changing the code to not return from the middle of the loop, running up to the end of the two compared data structures even if a mismatch has been detected. We hypothesized that our algorithm would not perform as well on this version of the code, and indeed, this posed a challenge, requiring 44 queries and 243 seconds on average to find the target for arrays of length just 2; the vulnerable version is crackable within 28 steps and 500 seconds on average up to arrays of length 6. The takeaway is that our algorithm can automatically find and exploit a vulnerability in code as well as demonstrate that a repair makes an exploit of the same kind infeasible.

\subsubsection{Hidden State Board Games}

\noindent \textbf{Mastermind.}
This is the classic board game in which a player tries to find a secret code consisting of 4 colored pegs where each peg can be one of 6 colors, by proposing their own 4-color code. The game responds by giving a number of red flags (the number of pegs in the correct positions with the correct color) and a number of white flags (the number of pegs with correct colors but in the incorrect positions)~\cite{Kooi05}. This game has garnered much attention with many publications describing solutions strategies. Donald Knuth gave the optimal solution that never needs more than 5 of steps. We see that our greedy information gain maximization approach achieves an average of 3.8 steps before determining the target color sequence. We achieve better than 5 because we are randomly choosing the target sequence and sometimes the search algorithm gets an `easy` code to break, whereas Knuth was working against a worst-case adversary. This game has also been shown to be NP-complete for $n$ pegs and 2 colors, again demonstrating the NP-completeness of our search problem statement in the more general case~\cite{mm-np-complete,Goodrich12,mmind}. 

\vskip 2mm

\noindent \textbf{Simple Mastermind.} 
This is a simplified version of Mastermind where only red flags are revealed. This makes the game slightly harder, since there is less information provided at each step.

\vskip 2mm

\noindent \textbf{Simple Battleship.} 
The player guesses two integers, coordinates in a grid of cells, attempting to sink a ship which takes up 3 vertically or horizontally adjacent cells. The game responds by saying whether the player has hit the hidden ship at those coordinates. This is a simplified version of the popular Battleship game in which there are 5 ships of different sizes~\cite{battleship}.




\subsubsection{Interactive User Preference Ranking}

\noindent \textbf{Movie Preferences.} 
An interactive system offers two movies at a time and the user says which of the two they prefer. The system hopes to discover the complete ranking of movies for the user. This is a version of the ranking via pairwise comparison problem~\cite{JamiesonN11} and is similar to an AI task in the existing literature in which one attempts to fully determine a customer's sushi preferences via pair-wise comparisons using queries from a data set~\cite{PuKS17}.


\subsubsection{Numeric and Array Searching}

\noindent \textbf{Low-High.} In this problem, the target is an unknown integer from a known range; the specification returns \texttt{low} if the query is less than the target, \texttt{equal} if it is equal to the target, and \texttt{high} otherwise.

\vskip 0.5mm
\noindent \textbf{Low-Middle-High.} 
This is the example game of Section~\ref{sec:intervalsearch}.

\vskip 0.5mm
\noindent \textbf{Unsorted Array.} In this problem, the target is simply an element in an array of unknown integers from a known range, and the specification returns \texttt{true} if $A[q^*] = t^*$ and returns \texttt{false} otherwise.

\vskip 0.5mm
\noindent \textbf{Sorted Array.} In this problem, the target is an element in a sorted array, and the specification returns \texttt{low} if $A[q^*] < t^*$, \texttt{equal} if $A[q^*] = t^*$, and \texttt{high} otherwise.

\subsubsection{Geometric Searching}

\noindent \textbf{Bounding Box 2 and 3 Dimensions.}
A common problem in computational geometry is to find an axis-parallel bounding box that tightly encloses a set of 
points~\cite{ORourke2004FindingME}. Within the constraints of our system, we modeled this by searching for a secret box within a grid. 


\vskip 0.5mm
\noindent \textbf{Pinpoint Via Half-Space Slicing.} In this problem, the goal is to find a point in by querying two axis aligned intervals. This is the 2D version of the Low-High problem. One $x$ axis and one $y$ axis point are chosen as queries, and the specification returns which of the 4 resulting subdivisions a target point is in.

\vskip 0.5mm
\noindent \textbf{3D 9-Way Split.} In this problem, the goal is to find a point in by querying three axis aligned intervals. This is the 3D version of the Low-High problem. One $x$ axis, one $y$ axis, and one $z$ axis point are chosen as queries, and the specification returns which of the 9 resulting subdivisions a target point is in. 



\subsection{Discussion of Experimental Results}

Our approach solves all of these search problems is reasonable amounts of time given only the source code of a specification of the search problem and access to the query evaluation function of an instantiation of the search problem. Overall, we observe that the bottleneck in our approach is the model counting and objective function optimization time, whereas symbolic execution is reasonably fast. The three most challenging games were Mastermind, Counterfeit Coin, and HorseRace, taking approximately 140 minutes, 34 minutes, and 10 minutes respectively. We observe that the time required for static analysis is always under 1 minute, except in the case of Horse Race, which is barely over a minute. The most expensive operations of the game solving phase are the model counting done by \textsc{Barvinok} and then maximizing the resulting entropy function. Further, we observe that the number of constraints is an important factor; search problems with a small number of observation constraints are typically more easily solved by our approach, even when the search space is large.

\section{Related Work}

Entropy maximization is a common technique for solving problems in various domains. Within the machine learning community, entropy maximization is the classic approach used in the ID3 algorithm \cite{DBLP:journals/ml/Quinlan86} and its variants to synthesize classification trees. The ID3 algorithm takes as input a labeled data set and associated features, and at each step the feature which splits the data into subgroups in such a way as to maximize information is chosen as the next test in the classification tree. In some sense, our approach may be considered a fully symbolic version of ID3 where the data to be classified is the target set and the features are instantiations of our outcome constraints with queries. Additionally, due to the huge tree sizes that would result from our approach, we synthesize only the path needed to discover an unknown target. Interesting future work would be to explore how our approach compares against ID3 on problems of searching within data sets or how ID3 might perform on problems similar to those that we solve. Another instance of entropy maximization for solving search problems is that of COBRA, which performs model counting by enumerating all possible queries and unknowns in puzzle games like Mastermind in order to find informative plays\cite{KlimosK15}. 

Using model counting and constraints derived from static analysis of code to compute probabilities of program behaviors is a common approach. For instance, earlier work presented at GANDALF 2018~\cite{DBLP:journals/corr/abs-1709-02092} used program constraints with the model counter LattE \cite{DeLoera20041273} to compute event probabilities in the context of game semantics. Probabilistic symbolic execution is itself an area of study that has been applied to reliability analysis of nondeterministic programs~\cite{DBLP:conf/icse/FilieriPV13}. Our approach differs in that we compute \textit{symbolic} probability functions over program inputs using a more powerful symbolic model counter, \textsc{Barvinok} rather than concrete probabilities using non-symbolic model counting approaches. Symbolic probabilities allow us to maximize the symbolic information gain function over program inputs (queries), in addition to enabling symbolic updates on knowledge about the target within the model of the interactive system. We note that symbolic information maximization has become a new technique in the domain of quantitative information flow analysis for  synthesizing side-channel vulnerabilities\cite{DBLP:journals/sigsoft/SahaEKBB19,DBLP:conf/eurosp/BangRB18,DBLP:conf/csfw/PhanBPMB17}. 

Finally, we observe that in all works we are aware of regarding synthesis of online adaptive solutions to various programmatically defined search problems from automatic game playing to software security analysis, suffer from scalability issues arising from the challenges of efficient model counting and static analysis~\cite{DBLP:journals/sigsoft/SahaEKBB19,DBLP:conf/eurosp/BangRB18,DBLP:conf/csfw/PhanBPMB17,DBLP:conf/icse/FilieriPV13}. Just as SMT solvers have increased the applicability of static analysis techniques like symbolic execution, we hope that advances in model counting will improve the scalability of quantitative symbolic analysis methods as well.

\section{Conclusion}

In this paper, we presented an approach to automatically solving search problems. Our meta-search algorithm takes a specification of a search problem, conveniently able to be written as a program in our domain specific language, and then, when provided access to an instantiation of that search problem, is able to automatically synthesis solution steps. Our approach works by performing symbolic execution on the specification of the search problem, using model counting to compute the probabilistic relationship between the search targets, queries, and outcomes, and maximizing an expected information gain function to adaptively synthesize queries which solve the problem online. We experimentally validated the effectiveness of our approach by implementing it and testing it on search problems from several domains. 

\nocite{*}
\bibliographystyle{eptcs}
\bibliography{biblio}
\end{document}

%% file: tables/data-tables.tex
\begin{table}[!htbp]
\footnotesize
\caption{Logical reasoning puzzles.}
\label{tab:results}
\centering
\resizebox{\textwidth}{!}{
\begin{tabular}{|l|l|l|l|l|l|l|l|l|}
\hline
\textcolor{white}{xxxxxxxxxxxxxx}&\textcolor{white}{xxxxxxxxxxxxxx}  & Average & Average  & Symbolic &  &  & & \\   
Problem ID & Details & Solve Time (s) & \# Rounds & Exec. Time (s) & $|\Psi|$  & $|\Phi|$ & $|Q|$ & $|T|$  \\ \hline \hline
Counterfeit Coin & 9 coins & 2072.795 & 3 &  7.146 &  54 & 3 & 8952 & 18  \\ \hline \hline
Horse Race & 5 horses 3 lanes & 599.923 &  3.5 &  68.132 &  60 & 6 & 125 & 120  \\ \hline
\end{tabular}} 

\caption{Hidden state board games.} 
\label{tab:results}
\centering
\resizebox{\textwidth}{!}{
\begin{tabular}{|l|l|l|l|l|l|l|l|l|}
\hline
\textcolor{white}{xxxxxxxxxxxxxx}&\textcolor{white}{xxxxxxxxxxxxxx}  & Average & Average  & Symbolic &  &  & & \\   
Problem ID & Details & Solve Time (s) & \# Rounds & Exec. Time (s) & $|\Psi|$  & $|\Phi|$ & $|Q|$ & $|T|$  \\ \hline \hline
Mastermind (MM)& 6 colors 1 peg & 1.019 & 3.6 &  0.194 &  2 & 2 & 6 & 6  \\ \hline
Mastermind (MM)& 6 colors 2 pegs & 3.431 & 3.3 &  0.995 &  7 & 5 & 36 & 36  \\ \hline
Mastermind (MM)& 6 colors 3 pegs & 38.498 & 3.1 &  5.666 &  34 & 9 & 216 & 216  \\ \hline
Mastermind (MM)& 6 colors 4 pegs & 8185.924 & 3.8 &  38.414 &  209 & 14 & 1296 & 1296  \\ \hline \hline
Simple MM & 6 colors 1 peg & 0.735 & 3.1 &  0.06 &  2 & 2 & 6 & 6  \\ \hline
Simple MM & 6 colors 2 pegs & 2.241 & 4.1 &  0.155 &  4 & 3 & 36 & 36  \\ \hline
Simple MM & 6 colors 3 pegs & 11.968 & 5.8 &  0.353 &  8 & 4 & 216 & 216  \\ \hline
Simple MM & 6 colors 4 pegs & 248.278 & 6.2 &  0.77 &  16 & 5 & 1296 & 1296  \\ \hline \hline
Battleship & 4x4 grid & 3.639 &  4.733 &  0.13 &  4 & 2 & 16 & 16  \\ \hline
Battleship & 8x8 grid & 17.984 &  11.333 &  0.115 &  4 & 2 & 64 & 96  \\ \hline
Battleship & 12x12 grid & 123.38 &  27.4 &  0.116 &  4 & 2 & 144 & 240  \\ \hline
\end{tabular}}


\caption{Security exploits. } 
\label{tab:results}
\centering
\resizebox{\textwidth}{!}{
\begin{tabular}{|l|l|l|l|l|l|l|l|l|}
\hline
\textcolor{white}{xxxxxxxxxxxxxx}&\textcolor{white}{xxxxxxx}  & Average & Average  & Symbolic &  &  & & \\   
Problem ID & Details & Solve Time (s) & \# Rounds & Exec. Time (s) & $|\Psi|$  & $|\Phi|$ & $|Q|$ & $|T|$  \\ \hline \hline

Password Checker & 2 digits & 5.484 &  10.55 &  0.08 &  3 & 3 & 100 & 100  \\ \hline
Password Checker & 4 digits & 27.235 &  15.75 &  0.15 &  5 & 5 & 10000 & 10000  \\ \hline
Password Checker & 6 digits & 501.127 &  28.55 &  0.246 &  7 & 7 & 1000000 & 1000000  \\ \hline \hline
Repaired Password Checker   &1 digit    &3.5897 &6.9    &0.0996 &2  &2  &10 & 10 \\ \hline
Repaired Password Checker   &2 digits   &243.0750   &44.1   &0.3059 &2  &2  & 100 & 100 \\ \hline

\end{tabular}}


\caption{Numeric and array searching.} 
\label{tab:results}
\centering
\resizebox{\textwidth}{!}{
\begin{tabular}{|l|l|l|l|l|l|l|l|l|}
\hline
\textcolor{white}{xxxxxxxxxxxxxx}& \textcolor{white}{xxxxxxxxxxxxxxxx} & Average & Average  & Symbolic &  &  & & \\   
Problem ID & Details & Solve Time (s) & \# Rounds & Exec. Time (s) & $|\Psi|$  & $|\Phi|$ & $|Q|$ & $|T|$  \\ \hline \hline
Low-High & from 1 to 10 & 0.65 & 2.5 &  0.03 &  3 & 3 & 10 & 10  \\ \hline
Low-High & from 1 to 100 & 1.386 & 5 &  0.03 &  3 & 3 & 100 & 100  \\ \hline
Low-High & from 1 to 1000 & 2.888 & 8.5 &  0.03 &  3 & 3 & 1000 & 1000  \\ \hline
Low-High & from 1 to 10000 & 6.405 & 12.3 &  0.03 &  3 & 3 & 10000 & 10000  \\ \hline
Low-High & from 1 to 100000 & 24.456 & 15.6 &  0.03 &  3 & 3 & 100000 & 100000  \\ \hline \hline

Low-Middle-High & from 1 to 10 & 2.727 & 5 &  0.033 &  3 & 3 & 81 & 45  \\ \hline
Low-Middle-High & from 1 to 50 & 24.643 & 10.1 &  0.032 &  3 & 3 & 2401 & 1225  \\ \hline
Low-Middle-High & from 1 to 100 & 89.158 & 11.9 &  0.032 &  3 & 3 & 9801 & 4950  \\ \hline \hline

Sorted Array      &  Length 8  & 5.7218  & 2.1  &2.1321 & 22 & 3 &  8  & 8 \\ \hline
Sorted Array      &  Length  16 & 24.4918 & 2.8  &9.6172 & 46 & 3 &  16 & 16 \\ \hline \hline
Unsorted Array  &  Length  8  & 9.7258  & 4.1  &2.0735 & 16 & 2 &  8  & 8 \\ \hline
Unsorted Array  &  Length  16 & 81.1663 & 8.4  &8.7484 & 32 & 2 &  16 & 16 \\ \hline


\end{tabular}}

\caption{Interactive Movie Ranking. } 
\label{tab:results}
\centering
\resizebox{\textwidth}{!}{
\begin{tabular}{|l|l|l|l|l|l|l|l|l|}
\hline
\textcolor{white}{xxxxxxxxxxxxxx}&\textcolor{white}{xxxxxxxxxxxxxx}  & Average & Average  & Symbolic &  &  & & \\   
Problem ID & Details & Solve Time (s) & \# Rounds & Exec. Time (s) & $|\Psi|$  & $|\Phi|$ & $|Q|$ & $|T|$  \\ \hline \hline
Movie Rank & 3 options & 2.725 &  2.6 &  0.679 &  6 & 2 & 9 & 6  \\ \hline
Movie Rank & 4 options & 16.559 &  4.533 &  1.325 &  12 & 2 & 16 & 24  \\ \hline
Movie Rank & 5 options & 188.955 &  6.933 &  2.381 &  20 & 2 & 25 & 120  \\ \hline
\end{tabular}}

\end{table}

\begin{table}[!htbp]

\caption{Geometric Searching. } 
\label{tab:results}
\centering
\resizebox{\textwidth}{!}{
\begin{tabular}{|l|l|l|l|l|l|l|l|l|}
\hline
\textcolor{white}{xxxxxxxxxxxxxx}&\textcolor{white}{xxxxxxxxxxxx}  & Average & Average  & Symbolic &  &  & & \\   
Problem ID & Details & Solve Time (s) & \# Rounds & Exec. Time (s) & $|\Psi|$  & $|\Phi|$ & $|Q|$ & $|T|$  \\ \hline \hline
2D Bounding Box & from 1 to 5 & 12.616 & 5.4 &  0.036 &  3 & 3 & 256 & 100  \\ \hline
2D Bounding Box & from 1 to 10 & 378.609 & 10.9 &  0.036 &  3 & 3 & 6561 & 2025  \\ \hline \hline

3D Bounding Box & from 1 to 3 & 16.622 &  4.2 &  0.04 &  3 & 3 & 729 & 27  \\ \hline
3D Bounding Box & from 1 to 4 & 160.525 &  6.9 &  0.04 &  3 & 3 & 729 & 216  \\ \hline \hline

Pinpoint & from 1 to 10 & 2.272 & 2.7 &  0.122 &  9 & 9 & 100 & 100  \\ \hline
Pinpoint & from 1 to 50 & 8.218 & 5 &  0.122 &  9 & 9 & 2500 & 2500  \\ \hline
Pinpoint & from 1 to 100 & 16.567 & 6 &  0.121 &  9 & 9 & 10000 & 10000  \\ \hline \hline

2D 9-Way Split & from 1 to 10 & 13.485 & 3 &  0.402 &  27 & 27 & 1000 & 1000  \\ \hline
2D 9-Way Split & from 1 to 50 & 101.692 & 4.1 &  0.404 &  27 & 27 & 27000 & 27000  \\ \hline
2D 9-Way Split & from 1 to 60 & 348.359 & 5.3 &  0.406 &  27 & 27 & 216000 & 216000  \\ \hline

\end{tabular}}
\end{table}

%% file: paper.bbl
\begin{thebibliography}{10}
\providecommand{\bibitemdeclare}[2]{}
\providecommand{\surnamestart}{}
\providecommand{\surnameend}{}
\providecommand{\urlprefix}{Available at }
\providecommand{\url}[1]{\texttt{#1}}
\providecommand{\href}[2]{\texttt{#2}}
\providecommand{\urlalt}[2]{\href{#1}{#2}}
\providecommand{\doi}[1]{doi:\urlalt{http://dx.doi.org/#1}{#1}}
\providecommand{\bibinfo}[2]{#2}

\bibitemdeclare{}{Xbox}
\bibitem{Xbox}
 (\bibinfo{year}{2007}): \emph{\bibinfo{title}{Xbox 360 Timing Attack}}.
\newblock
  \bibinfo{howpublished}{http://beta.ivc.no/wiki/index.php/Xbox_360_Timing_Attack}.

\bibitemdeclare{}{OAuth13}
\bibitem{OAuth13}
 (\bibinfo{year}{2013}): \emph{\bibinfo{title}{OAuth Protocol HMAC Byte Value
  Calculation Timing Disclosure Weakness}}.
\newblock \bibinfo{howpublished}{\url{https://osvdb.info/OSVDB-97562}}.

\bibitemdeclare{inproceedings}{ABB15}
\bibitem{ABB15}
\bibinfo{author}{Abdulbaki \surnamestart Aydin\surnameend},
  \bibinfo{author}{Lucas \surnamestart Bang\surnameend} \&
  \bibinfo{author}{Tevfik \surnamestart Bultan\surnameend}
  (\bibinfo{year}{2015}): \emph{\bibinfo{title}{Automata-Based Model Counting
  for String Constraints}}.
\newblock In: {\sl \bibinfo{booktitle}{Proceedings of the 27th International
  Conference on Computer Aided Verification (CAV)}},
  \doi{10.1007/978-3-540-85114-1_21}.

\bibitemdeclare{inproceedings}{BAP16}
\bibitem{BAP16}
\bibinfo{author}{Lucas \surnamestart Bang\surnameend},
  \bibinfo{author}{Abdulbaki \surnamestart Aydin\surnameend},
  \bibinfo{author}{Quoc-Sang \surnamestart Phan\surnameend},
  \bibinfo{author}{Corina~S. \surnamestart Pasareanu\surnameend} \&
  \bibinfo{author}{Tevfik \surnamestart Bultan\surnameend}
  (\bibinfo{year}{2016}): \emph{\bibinfo{title}{String Analysis for Side
  Channels with Segmented Oracles}}.
\newblock In: {\sl \bibinfo{booktitle}{Proc. of the 24th ACM SIGSOFT
  International Symp. on the Foundations of Software Engineering}},
  \doi{10.1145/2950290.2950362}.

\bibitemdeclare{inproceedings}{DBLP:conf/eurosp/BangRB18}
\bibitem{DBLP:conf/eurosp/BangRB18}
\bibinfo{author}{Lucas \surnamestart Bang\surnameend},
  \bibinfo{author}{Nicol{\'{a}}s \surnamestart Rosner\surnameend} \&
  \bibinfo{author}{Tevfik \surnamestart Bultan\surnameend}
  (\bibinfo{year}{2018}): \emph{\bibinfo{title}{Online Synthesis of Adaptive
  Side-Channel Attacks Based On Noisy Observations}}.
\newblock In: {\sl \bibinfo{booktitle}{2018 {IEEE} European Symposium on
  Security and Privacy, EuroS{\&}P 2018, London, United Kingdom, April 24-26,
  2018}}, \bibinfo{publisher}{{IEEE}}, pp. \bibinfo{pages}{307--322}.

\bibitemdeclare{article}{Barvinok:1994:PTA:187096.187093}
\bibitem{Barvinok:1994:PTA:187096.187093}
\bibinfo{author}{Alexander~I. \surnamestart Barvinok\surnameend}
  (\bibinfo{year}{1994}): \emph{\bibinfo{title}{{A polynomial time algorithm
  for counting integral points in polyhedra when the dimension is fixed}}}.
\newblock {\sl \bibinfo{journal}{Math. Oper. Res.}}
  \bibinfo{volume}{19}(\bibinfo{number}{4}), pp. \bibinfo{pages}{769--779},
  \doi{10.1287/moor.19.4.769}.

\bibitemdeclare{article}{accept-reject-sampling}
\bibitem{accept-reject-sampling}
\bibinfo{author}{George \surnamestart Casella\surnameend},
  \bibinfo{author}{Christian~P. \surnamestart Robert\surnameend} \&
  \bibinfo{author}{Martin~T. \surnamestart Wells\surnameend}
  (\bibinfo{year}{2004}): \emph{\bibinfo{title}{Generalized Accept-Reject
  Sampling Schemes}}.
\newblock {\sl \bibinfo{journal}{Lecture Notes-Monograph Series}}
  \bibinfo{volume}{45}, pp. \bibinfo{pages}{342--347},
  \doi{10.1214/lnms/1196285403}.
\newblock \urlprefix\url{http://www.jstor.org/stable/4356322}.

\bibitemdeclare{book}{Cover2006}
\bibitem{Cover2006}
\bibinfo{author}{Thomas~M. \surnamestart Cover\surnameend} \&
  \bibinfo{author}{Joy~A. \surnamestart Thomas\surnameend}
  (\bibinfo{year}{2006}): \emph{\bibinfo{title}{Elements of Information Theory
  (Wiley Series in Telecommunications and Signal Processing)}}.
\newblock \bibinfo{publisher}{Wiley-Interscience}.

\bibitemdeclare{inproceedings}{DeMoura:2008:ZES:1792734.1792766}
\bibitem{DeMoura:2008:ZES:1792734.1792766}
\bibinfo{author}{Leonardo \surnamestart De~Moura\surnameend} \&
  \bibinfo{author}{Nikolaj \surnamestart Bj{\o}rner\surnameend}
  (\bibinfo{year}{2008}): \emph{\bibinfo{title}{{Z3: an efficient SMT
  solver}}}.
\newblock In: {\sl \bibinfo{booktitle}{Proceedings of the 14th international
  conference on Tools and algorithms for the construction and analysis of
  systems}}, \bibinfo{series}{TACAS'08}, pp. \bibinfo{pages}{337--340}.

\bibitemdeclare{inproceedings}{DBLP:journals/corr/abs-1709-02092}
\bibitem{DBLP:journals/corr/abs-1709-02092}
\bibinfo{author}{Aleksandar~S. \surnamestart Dimovski\surnameend}
  (\bibinfo{year}{2017}): \emph{\bibinfo{title}{Probabilistic Analysis Based On
  Symbolic Game Semantics and Model Counting}}.
\newblock In \bibinfo{editor}{Patricia \surnamestart Bouyer\surnameend},
  \bibinfo{editor}{Andrea \surnamestart Orlandini\surnameend} \&
  \bibinfo{editor}{Pierluigi~San \surnamestart Pietro\surnameend}, editors:
  {\sl \bibinfo{booktitle}{Proceedings Eighth International Symposium on Games,
  Automata, Logics and Formal Verification, GandALF 2017, Roma, Italy, 20-22
  September 2017}}, {\sl \bibinfo{series}{{EPTCS}}} \bibinfo{volume}{256}, pp.
  \bibinfo{pages}{1--15}.

\bibitemdeclare{inproceedings}{DBLP:conf/icse/FilieriPV13}
\bibitem{DBLP:conf/icse/FilieriPV13}
\bibinfo{author}{Antonio \surnamestart Filieri\surnameend},
  \bibinfo{author}{Corina~S. \surnamestart Pasareanu\surnameend} \&
  \bibinfo{author}{Willem \surnamestart Visser\surnameend}
  (\bibinfo{year}{2013}): \emph{\bibinfo{title}{Reliability analysis in
  symbolic pathfinder}}.
\newblock In \bibinfo{editor}{David \surnamestart Notkin\surnameend},
  \bibinfo{editor}{Betty H.~C. \surnamestart Cheng\surnameend} \&
  \bibinfo{editor}{Klaus \surnamestart Pohl\surnameend}, editors: {\sl
  \bibinfo{booktitle}{35th International Conference on Software Engineering,
  {ICSE} '13, San Francisco, CA, USA, May 18-26, 2013}},
  \bibinfo{publisher}{{IEEE} Computer Society}, pp. \bibinfo{pages}{622--631}.

\bibitemdeclare{article}{Goodrich12}
\bibitem{Goodrich12}
\bibinfo{author}{Michael~T. \surnamestart Goodrich\surnameend}
  (\bibinfo{year}{2012}): \emph{\bibinfo{title}{Learning Character Strings via
  Mastermind Queries, With a Case Study Involving mtDNA}}.
\newblock {\sl \bibinfo{journal}{{IEEE} Trans. Information Theory}}
  \bibinfo{volume}{58}(\bibinfo{number}{11}), pp. \bibinfo{pages}{6726--6736},
  \doi{10.1109/TIT.2012.2208581}.

\bibitemdeclare{article}{HYAFIL197615}
\bibitem{HYAFIL197615}
\bibinfo{author}{Laurent \surnamestart Hyafil\surnameend} \&
  \bibinfo{author}{Ronald~L. \surnamestart Rivest\surnameend}
  (\bibinfo{year}{1976}): \emph{\bibinfo{title}{Constructing optimal binary
  decision trees is NP-complete}}.
\newblock {\sl \bibinfo{journal}{Information Processing Letters}}
  \bibinfo{volume}{5}(\bibinfo{number}{1}), pp. \bibinfo{pages}{15 -- 17},
  \doi{10.1016/0020-0190(76)90095-8}.

\bibitemdeclare{article}{battleship}
\bibitem{battleship}
\bibinfo{author}{W~\surnamestart J.~M.~Meuffels\surnameend} \&
  \bibinfo{author}{Dick \surnamestart den Hertog\surnameend}
  (\bibinfo{year}{2010}): \emph{\bibinfo{title}{Puzzle —Solving the
  Battleship Puzzle as an Integer Programming Problem}}.
\newblock {\sl \bibinfo{journal}{Journal of Financial Stability}}
  \bibinfo{volume}{10}, pp. \bibinfo{pages}{156--162}.

\bibitemdeclare{incollection}{JamiesonN11}
\bibitem{JamiesonN11}
\bibinfo{author}{Kevin~G \surnamestart Jamieson\surnameend} \&
  \bibinfo{author}{Robert \surnamestart Nowak\surnameend}
  (\bibinfo{year}{2011}): \emph{\bibinfo{title}{Active Ranking using Pairwise
  Comparisons}}.
\newblock In \bibinfo{editor}{J.~\surnamestart Shawe-Taylor\surnameend},
  \bibinfo{editor}{R.~S. \surnamestart Zemel\surnameend},
  \bibinfo{editor}{P.~L. \surnamestart Bartlett\surnameend},
  \bibinfo{editor}{F.~\surnamestart Pereira\surnameend} \&
  \bibinfo{editor}{K.~Q. \surnamestart Weinberger\surnameend}, editors: {\sl
  \bibinfo{booktitle}{Advances in Neural Information Processing Systems 24}},
  \bibinfo{publisher}{Curran Associates, Inc.}, pp.
  \bibinfo{pages}{2240--2248}.
\newblock
  \urlprefix\url{http://papers.nips.cc/paper/4427-active-ranking-using-pairwise-comparisons.pdf}.

\bibitemdeclare{article}{King:1976:SEP:360248.360252}
\bibitem{King:1976:SEP:360248.360252}
\bibinfo{author}{James~C. \surnamestart King\surnameend}
  (\bibinfo{year}{1976}): \emph{\bibinfo{title}{{Symbolic execution and program
  testing}}}.
\newblock {\sl \bibinfo{journal}{Commun. ACM}}
  \bibinfo{volume}{19}(\bibinfo{number}{7}), pp. \bibinfo{pages}{385--394},
  \doi{10.1145/360248.360252}.

\bibitemdeclare{inproceedings}{KlimosK15}
\bibitem{KlimosK15}
\bibinfo{author}{Miroslav \surnamestart Klimos\surnameend} \&
  \bibinfo{author}{Anton{\'{\i}}n \surnamestart Kucera\surnameend}
  (\bibinfo{year}{2015}): \emph{\bibinfo{title}{Cobra: {A} Tool for Solving
  General Deductive Games}}.
\newblock In: {\sl \bibinfo{booktitle}{Logic for Programming, Artificial
  Intelligence, and Reasoning - 20th International Conference, November
  24-28}}, pp. \bibinfo{pages}{31--47}, \doi{10.1016/j.tcs.2006.08.042}.

\bibitemdeclare{article}{Kooi05}
\bibitem{Kooi05}
\bibinfo{author}{Barteld~P. \surnamestart Kooi\surnameend}
  (\bibinfo{year}{2005}): \emph{\bibinfo{title}{Yet Another Mastermind
  Strategy}}.
\newblock {\sl \bibinfo{journal}{{ICGA} Journal}}
  \bibinfo{volume}{28}(\bibinfo{number}{1}), pp. \bibinfo{pages}{13--20},
  \doi{10.3233/ICG-2005-28105}.

\bibitemdeclare{inproceedings}{DBLP:conf/ccs/KopfB07}
\bibitem{DBLP:conf/ccs/KopfB07}
\bibinfo{author}{Boris \surnamestart K{\"{o}}pf\surnameend} \&
  \bibinfo{author}{David~A. \surnamestart Basin\surnameend}
  (\bibinfo{year}{2007}): \emph{\bibinfo{title}{An information-theoretic model
  for adaptive side-channel attacks}}.
\newblock In: {\sl \bibinfo{booktitle}{Proceedings of the 2007 {ACM} Conference
  on Computer and Communications Security, {CCS} 2007, Alexandria, Virginia,
  USA, October 28-31, 2007}}, pp. \bibinfo{pages}{286--296}.

\bibitemdeclare{}{Law09.2}
\bibitem{Law09.2}
\bibinfo{author}{Nate \surnamestart Lawson\surnameend} (\bibinfo{year}{2009}):
  \emph{\bibinfo{title}{Timing attack in Google Keyczar library}}.
\newblock
  \bibinfo{howpublished}{\url{https://rdist.root.org/2009/05/28/timing-attack-in-google-keyczar-library/}}.

\bibitemdeclare{article}{DeLoera20041273}
\bibitem{DeLoera20041273}
\bibinfo{author}{Jes\'us A.~De \surnamestart Loera\surnameend},
  \bibinfo{author}{Raymond \surnamestart Hemmecke\surnameend},
  \bibinfo{author}{Jeremiah \surnamestart Tauzer\surnameend} \&
  \bibinfo{author}{Ruriko \surnamestart Yoshida\surnameend}
  (\bibinfo{year}{2004}): \emph{\bibinfo{title}{Effective lattice point
  counting in rational convex polytopes}}.
\newblock {\sl \bibinfo{journal}{Journal of Symbolic Computation}}
  \bibinfo{volume}{38}(\bibinfo{number}{4}), pp. \bibinfo{pages}{1273 -- 1302},
  \doi{10.1016/j.jsc.2003.04.003}.
\newblock \bibinfo{note}{Symbolic Computation in Algebra and Geometry}.

\bibitemdeclare{inproceedings}{LSS14}
\bibitem{LSS14}
\bibinfo{author}{Loi \surnamestart Luu\surnameend}, \bibinfo{author}{Shweta
  \surnamestart Shinde\surnameend}, \bibinfo{author}{Prateek \surnamestart
  Saxena\surnameend} \& \bibinfo{author}{Brian \surnamestart Demsky\surnameend}
  (\bibinfo{year}{2014}): \emph{\bibinfo{title}{A model counter for constraints
  over unbounded strings}}.
\newblock In: {\sl \bibinfo{booktitle}{Proceedings of the {ACM} {SIGPLAN}
  Conference on Programming Language Design and Implementation (PLDI)}},
  p.~\bibinfo{pages}{57}, \doi{10.1109/SP.2009.8}.

\bibitemdeclare{article}{mmind}
\bibitem{mmind}
\bibinfo{author}{J~\surnamestart Maestro-Montojo\surnameend},
  \bibinfo{author}{Sancho \surnamestart Salcedo-Sanz\surnameend} \&
  \bibinfo{author}{Juan \surnamestart Merelo~Guervós\surnameend}
  (\bibinfo{year}{2014}): \emph{\bibinfo{title}{New solver and optimal
  anticipation strategies design based on evolutionary computation for the game
  of MasterMind}}.
\newblock {\sl \bibinfo{journal}{Evolutionary Intelligence}}
  \bibinfo{volume}{6}, \doi{10.1007/s12065-013-0099-6}.

\bibitemdeclare{}{Nel10}
\bibitem{Nel10}
\bibinfo{author}{Taylor \surnamestart Nelson\surnameend}
  (\bibinfo{year}{2010}): \emph{\bibinfo{title}{Widespread Timing
  Vulnerabilities in OpenID implementations}}.
\newblock
  \bibinfo{howpublished}{\url{http://lists.openid.net/pipermail/openid-security/2010-July/001156.html}}.

\bibitemdeclare{article}{ORourke2004FindingME}
\bibitem{ORourke2004FindingME}
\bibinfo{author}{Joseph \surnamestart O'Rourke\surnameend}
  (\bibinfo{year}{2004}): \emph{\bibinfo{title}{Finding minimal enclosing
  boxes}}.
\newblock {\sl \bibinfo{journal}{International Journal of Computer \&
  Information Sciences}} \bibinfo{volume}{14}, pp. \bibinfo{pages}{183--199},
  \doi{10.1007/BF00991005}.

\bibitemdeclare{inproceedings}{DBLP:conf/csfw/PhanBPMB17}
\bibitem{DBLP:conf/csfw/PhanBPMB17}
\bibinfo{author}{Quoc{-}Sang \surnamestart Phan\surnameend},
  \bibinfo{author}{Lucas \surnamestart Bang\surnameend},
  \bibinfo{author}{Corina~S. \surnamestart Pasareanu\surnameend},
  \bibinfo{author}{Pasquale \surnamestart Malacaria\surnameend} \&
  \bibinfo{author}{Tevfik \surnamestart Bultan\surnameend}
  (\bibinfo{year}{2017}): \emph{\bibinfo{title}{Synthesis of Adaptive
  Side-Channel Attacks}}.
\newblock In: {\sl \bibinfo{booktitle}{30th {IEEE} Computer Security
  Foundations Symposium, {CSF} 2017}}, \doi{10.1109/CSF.2017.8}.

\bibitemdeclare{inproceedings}{PuKS17}
\bibitem{PuKS17}
\bibinfo{author}{Yewen \surnamestart Pu\surnameend},
  \bibinfo{author}{Leslie~Pack \surnamestart Kaelbling\surnameend} \&
  \bibinfo{author}{Armando \surnamestart Solar{-}Lezama\surnameend}
  (\bibinfo{year}{2017}): \emph{\bibinfo{title}{Learning to Acquire
  Information}}.
\newblock In: {\sl \bibinfo{booktitle}{Proceedings of the Thirty-Third
  Conference on Uncertainty in Artificial Intelligence, {UAI} 2017, Sydney,
  Australia, August 11-15, 2017}}.

\bibitemdeclare{article}{DBLP:journals/ml/Quinlan86}
\bibitem{DBLP:journals/ml/Quinlan86}
\bibinfo{author}{J.~Ross \surnamestart Quinlan\surnameend}
  (\bibinfo{year}{1986}): \emph{\bibinfo{title}{Induction of Decision Trees}}.
\newblock {\sl \bibinfo{journal}{Mach. Learn.}}
  \bibinfo{volume}{1}(\bibinfo{number}{1}), pp. \bibinfo{pages}{81--106},
  \doi{10.1023/A:1022643204877}.

\bibitemdeclare{}{mm-np-complete}
\bibitem{mm-np-complete}
\bibinfo{author}{Anthony \surnamestart Rhodes\surnameend}
  (\bibinfo{year}{2019}): \emph{\bibinfo{title}{Search Algorithms for
  Mastermind}}.

\bibitemdeclare{article}{DBLP:journals/sigsoft/SahaEKBB19}
\bibitem{DBLP:journals/sigsoft/SahaEKBB19}
\bibinfo{author}{Seemanta \surnamestart Saha\surnameend},
  \bibinfo{author}{William \surnamestart Eiers\surnameend},
  \bibinfo{author}{Ismet~Burak \surnamestart Kadron\surnameend},
  \bibinfo{author}{Lucas \surnamestart Bang\surnameend} \&
  \bibinfo{author}{Tevfik \surnamestart Bultan\surnameend}
  (\bibinfo{year}{2019}): \emph{\bibinfo{title}{Incremental Attack Synthesis}}.
\newblock {\sl \bibinfo{journal}{{ACM} {SIGSOFT} Software Engineering Notes}}
  \bibinfo{volume}{44}(\bibinfo{number}{4}), p.~\bibinfo{pages}{16},
  \doi{10.1007/s10515-013-0122-2}.

\bibitemdeclare{article}{shannon48}
\bibitem{shannon48}
\bibinfo{author}{Claude \surnamestart Shannon\surnameend}
  (\bibinfo{year}{1948}): \emph{\bibinfo{title}{A Mathematical Theory of
  Communication}}.
\newblock {\sl \bibinfo{journal}{Bell System Technical Journal}}
  \bibinfo{volume}{27}, pp. \bibinfo{pages}{379--423, 623--656},
  \doi{10.1002/j.1538-7305.1948.tb00917.x}.

\bibitemdeclare{article}{coinproblem}
\bibitem{coinproblem}
\bibinfo{author}{Cedric A.~B. \surnamestart Smith\surnameend}
  (\bibinfo{year}{1947}): \emph{\bibinfo{title}{The Counterfeit Coin Problem}}.
\newblock {\sl \bibinfo{journal}{The Mathematical Gazette}}
  \bibinfo{volume}{31}(\bibinfo{number}{293}), pp. \bibinfo{pages}{31--39},
  \doi{10.2307/3608991}.
\newblock \urlprefix\url{http://www.jstor.org/stable/3608991}.

\bibitemdeclare{inproceedings}{Smi09}
\bibitem{Smi09}
\bibinfo{author}{Geoffrey \surnamestart Smith\surnameend}
  (\bibinfo{year}{2009}): \emph{\bibinfo{title}{On the Foundations of
  Quantitative Information Flow}}.
\newblock In: {\sl \bibinfo{booktitle}{Proceedings of the 12th International
  Conference on Foundations of Software Science and Computational Structures
  (FOSSACS)}}, \doi{10.1137/060651380}.

\bibitemdeclare{}{website:barvinok}
\bibitem{website:barvinok}
\bibinfo{author}{Sven \surnamestart Verdoolaege\surnameend}
  (\bibinfo{year}{2017}): \emph{\bibinfo{title}{Barvinok model counter}}.
\newblock \urlprefix\url{http://barvinok.gforge.inria.fr//}.

\end{thebibliography}
